\documentclass[twocolumn]{aastex62}
\shorttitle{FRBs from interacting BNS systems}
\shortauthors{Zhang}
\begin{document}


\title{Fast Radio Bursts from Interacting Binary Neutron Star Systems}
%\title{Is the repeating FRB 121102 representative of Fast Radio Bursts?}
%\title{Do all Fast Radio Bursts repeat?}

%% Use \author, \affil, and the \and command to format author and affiliation 
%% information.  If done correctly the peer review system will be able to
%% automatically put the author and affiliation information from the manuscript
%% and save the corresponding author the trouble of entering it by hand.
%%
%% The \affil should be used to document primary affiliations and the
%% \altaffil should be used for secondary affiliations, titles, or email.

%% Authors with the same affiliation can be grouped in a single
\author{Bing Zhang }
\affil{Department of Physics and Astronomy, University of Nevada, Las Vegas, Las Vegas, NV 89154,
zhang@physics.unlv.edu}
\affil{Center for Gravitational Physics, Yukawa Institute for Theoretical Physics, Kyoto University, Kyoto, 606-8502, Japan}

\begin{abstract}
Recent observations of repeating fast radio bursts (FRBs) suggest that some FRBs reside in an environment consistent with that of binary neutron star (BNS) mergers. The bursting rate for repeaters could be very high and the emission site is likely from a magnetosphere.  We discuss a hypothesis of producing abundant repeating FRBs in BNS systems. Decades to centuries before a BNS system coalesces, the magnetospheres of the two neutron stars start to interact relentlessly. Abrupt magnetic reconnection accelerates particles, which emit coherent radio waves in bunches via curvature radiation.  FRBs are detected as these bright radiation beams point towards Earth. This model predicts quasi-periodicity of the bursts at the rotation periods of the two merging neutron stars (tens of milliseconds and seconds, respectively) as well as the period of orbital motion (of the order of 100 s).  The bursting activities are expected to elevate with time as the two neutron stars get closer. The repeating FRB sources should be gravitational wave (GW) sources for space-borne detectors such as Laser Interferometer Space Antenna (LISA), and eventually could be detected by ground-based detectors when the two neutron stars coalesce.
 \end{abstract}

\keywords{radio continuum: general -- stars: neutron -- binaries: general -- magnetic reconnection -- gravitational waves}

\section{Introduction} \label{sec:intro}

Despite rapid progress in the field of fast radio bursts (FRBs) \citep{lorimer07,thornton13}, the origin of these bursts is still mysterious \citep{petroff19,cordes19}. Recent observational progress suggests that repeaters are common \citep{spitler16,scholz16,chime-2ndrepeater,chime-repeaters,kumar19} and that the localized FRBs are harbored in diverse types of host galaxies \citep{tendulkar17,bannister19,ravi19}. The following observational properties of repeating FRBs are noticeable, which pose important constraints on any successful source model:
\begin{itemize}
\item The rate of repeating bursts could be very high at least for some sources, e.g. FRB 121102 \citep{law17,zhangy18,li20} and FRB 180301 \citep{luo20}.  This may suggest that the production of bursts is energetically inexpensive\footnote{The large number of bursts greatly raises the demands in most models, both intrinsic (e.g. the magnetar models that invoke starquakes, \citealt{wang18}, or spontaneous magnetic reconfigurations, \citealt{katz18}) and extrinsic (e.g. the comet/asteroid-hitting-neutron-star model, \citealt{dai16,smallwood19}) ones, because each burst requires a fresh trigger, which may not be easily realized in these models.}. 
\item The repeating activities seem not decline with time during the timescale of a few years, as observed in FRB 121102\footnote{Popular spindown-powered or magnetically-powered young magnetar models \citep{murase16,metzger17,beloborodov17} predict that the level of burst activities should die out with time. {In principle, the observational time for FRB 121102 may still not be long enough to test this prediction yet. In any case, young magnetars should have already entered the $\dot E \propto t^{-2}$ phase ($\dot E$ is the spindown power of the magnetar) in the timescale of a decade. Long-term monitoring of FRB 121102 and other active FRBs would be essential to test this prediction.}}.
\item The dispersion measure (DM) of FRB 121102 does not evolve during the period of multiple years\footnote{This poses constraints on the models invoking an expanding supernova remnant shell \citep{metzger17,yangzhang17,piro18}.}. The rotation measure (RM) of FRBs, on the other hand, show significant secular \citep{michilli18} and short-term variations \citep{luo20}. This suggests a dynamical magneto-environment in the vicinity of the FRB sources. 
\item Whereas the host galaxy of FRB 121102 is a dwarf star-forming galaxy similar to those of long gamma-ray bursts (GRBs) and superluminous supernovae \citep{chatterjee17,marcote17,tendulkar17,nicholl17}, most other FRB hosts are old, massive galaxies similar to the Milky Way, with the FRB source location having an offset from the center of the host \citep{bannister19,ravi19,marcote20}. These properties are consistent with those of short GRBs that are believed to have a binary neutron star (BNS) merger origin. A connection between FRBs and BNS mergers is tempting\footnote{Some one-off FRB models invoking catastrophic events during or shortly after BNS mergers have been proposed \cite[e.g.][]{totani13,zhang14,wang16}. However, the event rate density of BNS mergers \citep{GW170817} is much smaller than that of FRBs \citep{luo19}.  \cite{margalit19} (see also \citealt{wang20}) proposed that some BNS mergers leave behind massive, stable, rapidly-spinning magnetars, which may power repeating FRBs. In order to account for the prevalence of the short-GRB-like hosts of FRBs, the fraction of stable neutron star merger remnants should be high \citep{gao16}, which is inconsistent with the claimed low ($<3\%$) fraction assuming that the merger product of GW170817 is a black hole \citep{margalit19b}.}.
\item The durations of the repeating FRBs are relatively long and show complicated temporal features \citep{chime-repeaters,luo20}, which are consistent with an underlying complicated magnetospheric structure. A subpulse down-drifting pattern seems common in at least some bursts \citep{hessels19,chime-2ndrepeater}, which is consistent with coherent curvature radiation from the open field line regions of neutron star magnetospheres \citep{wang19}.
\item Observations of FRB 180301 repeating bursts show variation of the polarization angle during each burst, suggesting a magnetospheric origin of the bursts \citep{luo20}. These variations show diverse patterns that are inconsistent with the simple rotation-vector model for radio pulsars, suggesting a more complicated magnetic geometry in the emission region.
\end{itemize}

Here we propose a hypothetical scenario to interpret all these observational features. This scenario borrows the idea of our previous interacting model for repeating FRBs \citep{zhang17,zhang18b}, but differ from it by invoking interacting BNS systems.  FRBs are envisaged to be sporadically produced for decades to centuries before the merger of a BNS system, as the magnetospheres of the two neutron stars interact relentlessly. {In the literature, some authors \citep{piro12,wang16,wang18b,metzger16,most20} have studied magnetosphere interactions of merging BNSs as well as their possible connection with FRBs. Other FRB models involving BH-NS mergers \citep[e.g.][]{mcwilliams11,mingarelli15,zhang19,dai19} or BH-BH mergers \citep{zhang16,liebling16,liu16,fraschetti18} have been also discussed. However, these studies focused on the epoch right before the merger, so that the generated FRBs are one-off events. Those models are very different from the repeating FRB model proposed in this Letter.}

\section{The model}
\subsection{Energy budget}

Repeating FRBs seem to have lower luminosities than apparently non-repeating ones, with a typical isotropic value of a few $10^{41} \ {\rm erg \ s^{-1}}$ \citep{luo20}. Given that the typical duration of repeating FRBs is a few ms, the isotropic energy of each burst can be estimated as $E_{\rm iso} \sim 10^{39}$ erg. The average isotropic-equivalent FRB production power from the source may be estimated as
\begin{equation}
    \bar L_{\rm FRB,iso} \sim \dot N E_{\rm iso} = (10^{42} \ {\rm erg \ yr^{-1}}) \dot N_3 E_{\rm iso,39},
\end{equation}
where $\dot N = 10^3 \ {\rm yr^{-1}} \dot N_3$ is the bursting rate (beaming toward Earth) per year from a particular source. For an FRB source lasting for a duration $\tau = (10^2 {\rm yr}) \ \tau_2$, the total isotropic-equivalent energy output in FRBs is
\begin{equation}
    E_{\rm FRB,iso} = \bar L_{\rm FRB,iso} \tau = (10^{44} \ {\rm erg}) \dot N_3 E_{\rm iso,39} \tau_2.
\label{eq:E_FRB_iso}
\end{equation}
{When beaming is considered, this energy budget is reduced. Let us assume that each FRB has a beaming angle of $\delta \Omega \ll 4\pi$ (e.g. of the order of $\sim \pi\gamma^{-2}$ in our scenario, where $\gamma$ is the characteristic Lorentz factor of electrons in the bunch), and that the bulk of FRBs are concentrated in a solid angle of $\Delta \Omega < 4\pi$ (which is expected for the interacting model discussed here). The true energy of each burst is smaller by a factor $\delta\Omega/4\pi$, and the total number of bursts is increased by a factor $\Delta\Omega/\delta\Omega$. As a result, the true FRB energy budget is
\begin{equation}
    E_{\rm FRB} = f_b E_{\rm FRB,iso}  = (10^{43} \ {\rm erg}) f_{b,-1} \dot N_3 E_{\rm iso,39} \tau_2,
\label{eq:E_FRB}
\end{equation}
where $f_b \equiv \Delta\Omega / 4\pi$. This energy should be the minimum energy budget in the system.
}

To estimate the total energy budget in the BNS system, we take 
the double pulsar system PSR J0737-3039A/B \citep{kramer08} as the nominal system. This system is the only BNS system whose both members have measured spin parameters. For reference, we list the relevant parameters of the two pulsars in Table 1. One can see that relatively speaking PSR A has a shorter period ($P$), lower polar cap magnetic field ($B_p$), but a higher spindown power ($\dot E$) than PSR B. We do not list the current orbital parameters of the system, since we envisage a much later stage of the evolution as the magnetospheres of the two pulsars interact. We do not assume longer periods of the two pulsars than observed in PSR J0737-3039A/B, since the observed BNS merger systems by LIGO/Virgo have  shorter lifetimes than Galactic BNS systems in order to merge within the Hubble time. In the following, we normalize the parameters of the two pulsars as the measured values from the PSR J0737-3039A/B system \citep{kramer08}, i.e. $P_{\rm A} = (0.0227 \ {\rm s}) \ p_{\rm A}$, $B_{\rm p,A} = (1.3\times 10^{10} \ {\rm G}) \ b_{\rm p,A}$,  $R_{\rm LC,A} = 1.1 \times 10^8 \ {\rm cm} \ r_{\rm LC,A}$, $\dot E_{\rm A} = 5.7 \times 10^{33} \ {\rm erg \ s^{-1}} \ \dot e_{\rm A}$; $P_{\rm B} = (2.77 \ {\rm s}) \ p_{\rm B}$, $B_{\rm p,B} = (3.2\times 10^{12} \ {\rm G}) \ b_{\rm p,B}$,  $R_{\rm LC,B} = 1.3\times 10^{10} \ {\rm cm} \ r_{\rm LC,B}$, $\dot E_{\rm A} = 1.6 \times 10^{30} \ {\rm erg \ s^{-1}} \ \dot e_{\rm B}$.

\begin{table}
\begin{center}
\caption{Parameters of the double pulsar system PSR J0737-3039A/B \citep{kramer08}.}
\begin{tabular}{ccc}
\tableline\tableline
Parameters & PSR J0737-3039A & PSR J0737-3039B \\
\tableline
$P$ & 22.7 \ {\rm ms} & 2.77 \ {\rm s} \\
$\dot P$ & $1.7\times 10^{-18}$ & $0.88 \times 10^{-15}$ \\
$B_p$ & $1.3\times 10^{10}$ G & $3.2\times 10^{12}$ G \\
$R_{\rm LC}$ & $1.1\times 10^8$ cm & $1.3\times 10^{10}$ cm \\
$\dot E$ & $5.7\times 10^{33} \ {\rm erg \ s^{-1}}$ & $1.6\times 10^{30} \ {\rm erg \ s^{-1}}$ \\
$M$ & 1.337(5)$M_\odot$ & 1.250(5) $M_\odot$ \\
\tableline
\end{tabular}
\end{center}
\label{tab:dPSR}
\end{table}

The ultimate energy budget in the system includes the rotation energies of the two pulsars:
\begin{eqnarray}
 E_{\rm rot,A} & = & \frac{1}{2} I \Omega_{\rm A}^2 = (3.8\times 10^{49} \ {\rm erg}) \ I_{45} p_{\rm A}^{-2}, \label{eq:ErotA} \\
 E_{\rm rot,B} & = & \frac{1}{2} I \Omega_{\rm B}^2 = (2.6\times 10^{45} \ {\rm erg}) \ I_{45} p_{\rm B}^{-2}, \label{eq:ErotB}
\end{eqnarray}
as well as the orbital gravitational energy releasable until coalescence
\begin{equation}
    E_{\rm orb} = \frac {G M_1 M_2} {2R} = (2.6 \times 10^{53} \ {\rm erg}) \ M_{1.4}^2 R_6^{-1},
\label{eq:Eorb}
\end{equation}
where $I=10^{45} \ {\rm g \ cm^2} \ I_{45}$ is the moment of inertia of the neutron stars, $M_1 = M_2 = (1.4 M_\odot) M_{1.4}$ and $R = 10^6 \ {\rm cm} \ R_6$ is the radius of the neutron stars. Several remarks should be made: (1) The magnetic energies of the two pulsars are $E_{\rm B,A} = (1/6) B_{\rm A}^2 R^3 = (2.8 \times 10^{37} \ {\rm erg}) \ b_{\rm A}^2$ and $E_{\rm B,B} = (1/6) B_{\rm B}^2 R^3 = (1.7 \times 10^{42} \ {\rm erg}) \ b_{\rm B}^2$, respectively. These energies (especially that of PSR B) can be directly dissipated to power FRB emission. However, after dissipation, it is likely that the fields would be replenished from the rotation energies of the neutron stars (by analogy with the magnetic cycle of the Sun). So we list the rotation energies of the two neutron stars (rather than their magnetic energies) as the ultimate energy sources. (2) Based on the face values of the spindown rates of the two pulsars, the usable spin energy during the period of $\tau$ is only $\sim \dot E \tau$, which is $(1.8 \times 10^{43} \ {\rm erg}) \ \dot e_{\rm A} \tau_2$ and $(5.0 \times 10^{39} \ {\rm erg}) \ \dot e_{\rm B} \tau_2$ for PSR A and B, respectively. This is barely enough to meet the repeating FRB energy budget unless $\Delta\Omega \ll 1$ or $\dot e_{\rm A} \gg 1$. However, due to the close interactions between the magnetospheres of the two pulsars, additional braking is possible to tap the spin energies of both pulsars, which are limited by Eqs.(\ref{eq:ErotA}) and (\ref{eq:ErotB}) and are more than enough to power the observed FRBs. (3) The majority of the orbital energy (Eq.(\ref{eq:Eorb})) is carried away by gravitational waves. However, it is likely that a small fraction of the orbital energy is dissipated due to the interaction between the two magnetospheres \citep[e.g.][]{palenzuela13a,palenzuela13b,carrasco20}. If this fraction is greater than $10^{-9}$, it would also provide another relevant energy budget to power repeating FRBs\footnote{According to Eqs.(20) and (21) of \cite{lai12}, the Alfv\'en drag energy dissipation rate is much smaller than $10^{-9}$ for the nominal parameters adopted in this Letter, so that the orbital gravitational energy may not contribute significantly to power FRBs.}.

\subsection{Timescales}

There are several characteristic timescales in a BNS system. The first two are the rotation periods of the two pulsars, which are typically of the order of tens of milliseconds and seconds, respectively. Since the triggers of bursts depend on the complicated magnetic configurations in the system, the arrival times of the detected bursts would not follow the same rotation phase as in radio pulsars so that no strict periodicity is  expected\footnote{In the case of rotating radio transients (RRATs) \citep{mclaughlin06}, even though pulses are sporadically emitted, one can still easily identify their periods since the RRAT magnetospheres are not subject to distortions due to interactions.}. In any case, the imprints of the two spin periods may still exist, probably in the form of some quasi-periodic features in the burst arrival times. This prediction can be tested with future repeating FRB data.

The third timescale is the orbital period, which we estimate below. Since PSR A is much more energetic than PSR B, its pulsar wind will significantly distort the magnetosphere of the latter. The pressure balance at the interaction front may be written as
\begin{equation}
    \frac{\dot E_{\rm A}}{4\pi r_{\rm A}^2 c} = \frac{B_{\rm p,B}^2}{8\pi} \left(\frac{R}{r_{\rm B}}\right)^6,
\label{eq:p-balance}
\end{equation}
where $r_{\rm A}$ and $r_{\rm B}$ are distances of the interaction front from PSRs A and B, respectively, and a dipolar magnetic field configuration has been assumed for B's magnetosphere. Significant interaction occurs as the separation between the two pulsars is comparable to the size of the distorted B's magnetosphere. This corresponds to $r_{\rm A} \sim r_{\rm B}$. Solving Eq.(\ref{eq:p-balance}), one gets the separation between the two pulsars
\begin{equation}
    a \sim 2 r_{\rm A} = 2 \left(\frac{B_{\rm p,B}^2 R^6 c}{2 \dot E_{\rm A}}\right)^{1/4} \simeq 4.5\times 10^9 \ {\rm cm} \ b_{\rm p,B}^{1/2} \dot e_{\rm A}^{-1/4}.
\end{equation}
Assuming a circular orbit and again  $M_1 = M_2 = 1.4 M_\odot$,  one can derive the orbital period of the system
\begin{equation}
 P_{\rm orb} = \left(\frac{4 \pi^2 a^3}{G M} \right)^{1/2}
 \simeq 100 \ {\rm s} \ \left(\frac{a}{4.5\times 10^9 \ {\rm cm}}\right)^{3/2}
 M_{2.8}^{-1/2}, 
\end{equation}
where $M=M_1+M_2  = (2.8 M_\odot) M_{2.8}$ is the total mass of the system. 
It would be interesting to look for a characteristic timescale of this order in the repeating FRB data.

The fourth timescale is the time towards the coalescence, which can be estimated as
\begin{equation}
    \tau \simeq 500 \ {\rm yr} \left(\frac{P_{\rm orb}}{100 \ {\rm s}}\right)^{8/3} \left(\frac{2.8 M_\odot}{M}\right)^{2/3} \left(\frac{0.7 M_\odot}{\mu}\right),
\end{equation}
where $\mu = M_1M_2/M$ is the reduced mass of the binary system. This is the typical lifetime of a repeating FRB source. Noticing the sensitive dependence (index 8/3) on $P_{\rm orb}$, this timescale may range from decades to centuries when a range of PSR parameters $(p_{\rm A}, b_{\rm A}, p_{\rm B}, b_{\rm B})$ are considered.

\subsection{Production of FRBs}

Within this model, the FRBs are conjectured to be produced during sudden reconnection of magnetic field lines. The magnetic geometry of an interacting BNS is complicated. It is difficult to provide concrete predictions on when a burst could be generated. Nonetheless, one may imagine that for certain configurations, magnetic field lines with opposite polarities from the two pulsars would encounter and reconnect, leading to active bursting episodes. The quiescent states correspond to the epochs when the magnetic configurations are not favorable for reconnection, or when the depleted magnetic fields are being replenished. Dedicated numerical simulations may reveal the complicated interaction processes in such systems.

The FRB radiation mechanism is very likely bunching coherent curvature radiation \citep{katz16,kumar17,yangzhang18}\footnote{Alternatively, coherent radio emission may be generated directly from reconnection-driven fast magnetosonic waves \citep{lyubarsky20}. }. In particular, \cite{yangzhang18} showed that a sudden deviation of the electric charge density from the nominal value (e.g. the Goldreich-Julian value, \citealt{goldreich69}) would induce coherent bunching curvature radiation. Such a condition is readily satisfied in a dynamically interacting system. The emission configuration is very similar to that of the ``cosmic comb'' model \citep{zhang17}, so that the estimate of the characteristic frequency and duration from that model can be directly applied, i.e.
\begin{equation}
    \nu = \frac{3}{4\pi} \frac{c}{\rho}\gamma_e^3 \simeq (7.2\times 10^8 \ {\rm Hz}) \ \rho_{10}^{-1} \gamma_{e,3}^3,
\end{equation}
and 
\begin{equation}
    \Delta t \sim \frac{a}{v \gamma_e} \simeq (3.3 {\rm ms}) \ a_{10} \beta_{-1}^{-1} \gamma_{e,3}^{-1}.
\end{equation}
Here $\rho$ is the curvature radius, which is comparable to the separation $a$ between the two pulsars, $\gamma_e \sim 10^3 \gamma_{e,3}$ is the typical Lorentz factor of the electrons accelerated from the reconnection regions, and $\beta = v/c$ is the dimensionless field-line-sweeping velocity of the emission region, which is normalized to $\sim 0.1$ of the light cylinder radius of PSR B.

The predicted FRB luminosity depends on the intrinsic properties of each reconnection and how the beamed emission intersects with the line of sight. Only very energetic events or the events whose beam squarely sweeps across earth would produce rare, extremely bright FRBs. Most FRBs should be less luminous and would follow a power law distribution in the apparent luminosity with the concrete power law index depending on  model details. The reconnection-injected particles likely slide along field lines after sychrotron cooling. Coherent bunching curvature radiation is preferentially produced in open field line regions \citep{yangzhang18}, which may interpret the observed sub-pulse down-drifting patterns in some bursts \citep{wang19}. There should be an associated high-energy emission for each burst with a luminosity $L_{\rm HE} \sim 10^{43} \ {\rm erg \ s^{-1}} \ \eta_{-2}$, which depends on the radio efficiency parameter $\eta$ (normalized to $10^{-2}$)\footnote{Since the true energy of each FRB is about $10^{39} \ {\rm erg} \cdot \pi \cdot 10^{-6} \sim 3\times 10^{33} \ {\rm erg}$ for our nominal parameters, $\eta = 10^{-2}$ corresponds to the case that each FRB consumes $\sim 3\times 10^{35} \ {\rm erg}$ magnetic energy from PSR B, which is about $\sim 10^{-7}$ of the total magnetic energy available in PSR B. The magnetic energy density decreases with radius sharply. We believe that this estimate of efficiency is reasonable for typical bursts when interactions just started. At the later epochs of the inspiral phase (which lasts for a shorter duration), more energy is available in each reconnection. This would give rise to  brighter FRBs and brighter high-energy counterparts. According to this model, rare, bright repeating FRBs may exist during the later phase of the inspiral.}. A millisecond-duration X-ray or $\gamma$-ray burst with such a luminosity at a typical FRB distance is way below the sensitivity of the current high-energy detectors.

\subsection{DM, RM, and polarization properties}

Unlike young supernova remnants, BNS  are old systems not surrounded by a matter shell in the immediate environment. As a result, one does not expect a significant contribution to DM from the vicinity of the bursting source. This is consistent with the observations of the FRBs residing in BNS-like environments \citep{bannister19,ravi19,marcote20}. Since the variation of other DM components is very small \citep{yangzhang17}, one does not expect DM evolution in this scenario. This is consistent with the data of repeating FRBs so far \citep{spitler16,law17,luo20}. Observations of significant DM variations would disfavor this model\footnote{There was a report that DM of FRB 121102 might show a slight increase  \citep{josephy19}. If this is confirmed, it would support the model invoking a supernova remnant in the coasting phase \citep{yangzhang17,piro18}.}.

The RM, on the other hand, is usually dominated by the immediate environment of the source where magnetic field strength is high. In a dynamically interacting system, one expects a complicated magnetic structure surrounding the system, so that RM, which depends on the integral of the parallel component of the magnetic field, can vary significantly within a short period of time. The evolution is also expected not to be monotonic. This is consistent with the observations of FRB 180301 \citep{luo20}. A BNS system is not expected to produce extremely large RMs. Within this model, the BNS system powering FRB 121102 is located near a supermassive black hole, which gives rise to the abnormally large RM for that source. The secular RM variation could be due to the orbital motion of the system around the black hole \citep{zhang18b}.

Coherent curvature radiation is intrinsically linearly polarized. Pulsar radio emission shows high linear polarization degrees and a signature sweeping pattern of the polarization angle in the form of ``S'' or inverse ``S'' patterns. This has been well-interpreted within the rotating vector model \citep{rv69} where coherent emission originates from the open field line region of an isolated rotating neutron star. 
In an interacting BNS system, the magnetosphere structure is much more complicated. One would expect the deviation from the simple rotating vector model and diverse polarization angle evolution patterns. These are consistent with the observations of the repeating bursts detected from FRB 180301 \citep{luo20}.  Under certain conditions (e.g. similar to the cosmic comb configuration as discussed in \citealt{zhang18b}), the emission region may be on nearly straight field lines. As the emission beam sweeps the line of sight, the polarization angle would not show significant evolution within single bursts. The absolute values of the polarization angles should vary among different bursts. Such a feature is not inconsistent with the observations of FRB 121102 \citep{michilli18}.

\subsection{Event rate density}

The event rate density of BNS mergers is estimated as ${\cal R}_{\rm BNS} \sim 1.5^{+3.2}_{-1.2}\times 10^3 \ {\rm Gpc^{-3} \ yr^{-1}}$ from the GW170817 detection \citep{GW170817}. That of FRBs above $10^{42} \ {\rm erg \ s^{-1}}$ is ${\cal R}_{\rm FRB}(>10^{42} \ {\rm erg \ s^{-1}}) = 3.5^{+5.7}_{-2.4} \times 10^4 \ {\rm Gpc^{-3} \ yr^{-1}}$ \citep{luo19}, which is $\sim 20$ times higher. Repeating FRBs typically have luminosities below $10^{42} \ {\rm erg \ s^{-1}}$ \citep{luo20}. Including these faint bursts, the FRB event rate density  may be boosted by another $\sim (2-3)$ orders of magnitude. If each BNS merger system produces $10^5$ bursts during its lifetime (our nominal value), one would over-produce FRBs by about (1-2) orders of magnitude. This suggests that either the average total number of bursts produced in BNS systems is lower (i.e. FRB 121102 is abnormally active, e.g. \citealt{palaniswamy18,caleb19}) or some interacting systems cannot produce FRBs because of their unfavorable pulsar parameters. 

\section{Summary and predictions}

We proposed a new hypothesis for repeating FRBs in this paper. BNS systems decades to centuries before merging would render the magnetospheres of the two neutron stars relentlessly interacting with each other. Abrupt magnetic reconnection during these interactions would inject particles which produce FRBs via coherent bunching curvature radiation in the magnetospheres of the neutron stars. 

This model could in principle interpret the following interesting observational facts (as listed in Section 1): the high event rate, no evidence of the decline of the burst rate in FRB 121102, non-evolution of DM in FRB 121102, rapid evolution of RM in FRB 180301, complicated temporal structure and polarization angle swing in the bursts of FRB 180301, sub-pulse down-drifting as observed in many bursts, as well as the host galaxy properties of a growing number of FRBs that show short-GRB-like (BNS merger) environments.  

An immediate prediction of this model is that repeating FRB sources are gravitational wave (GW) sources whose frequencies ($\sim 10^{-2}$ Hz) fall into the range of the space-borne GW detectors such as LISA \citep{LISA}, TaiJi \citep{Taiji} and TianQin \citep{Tianqin}. Observations of some nearby FRB sources within the horizons of these gravitational detectors in 2030s would be a direct test of this model. These sources would eventually be detected by ground-based kHz GW detectors such as the successors of LIGO/Virgo detectors, when the BNS coalesce decades to centuries later.

This model also predicts that the bursting activities of the repeating FRB sources (such as FRB 121102 and FRB 180301) should not decline, and would elevate with time as the two neutron stars get closer and closer. Observations of enhanced activities from these sources could be an indirect support to the model.

Finally, during the refereeing process of this paper, a $\sim 16$-day period was announced for the CHIME repeating source FRB 180916.J0158+65 \citep{chime-period}. This period is best understood as the orbital period of a binary system, but is too long compared with the orbital period predicted in this paper ($\sim 100$ s). That event may be interpreted within the context of cosmic-comb-induced binary interaction models \citep{ioka20}. 

\acknowledgments 
I thank Z. G. Dai, K. Ioka, D. Lai, K. J. Lee, C. Thompson, and X.-F. Wu for interesting discussions and an anonymous referee for comments. I am also grateful to UNLV for granting me a sabbatical leave and to YITP for the hospitality during my stay when this work was finished.

%\bibliography{FRB}

\end{document}